\documentstyle[twocolumn,prb,aps]{revtex}

\begin{document}

\relax
\draft \preprint{RU/1-gg}

\title{Chaotic dynamics, fluctuations, nonequilibrium ensembles.}

\author{Giovanni Gallavotti}
\address{Fisica, Universit\'{a} di Roma, ``La Sapienza'', 00185
Roma, Italia}

\date{\today}
\maketitle
\begin{abstract}
A review of some recent results and ideas about the
expected behaviour of large chaotic systems and fluids.
\end{abstract}



\begin{section}{Ergodic hypothesis}
\narrowtext

Giving up a detailed description of microscopic motion led to a
statistical theory of macroscopic systems and to a deep understanding
of their equilibrium properties.  At the same time a far less
successful (even for steady states theory) approach to nonequilibrium
systems began.

It is clear today, as it was already to Boltzmann and many
others, that some of the assumptions and guiding ideas used in
building up the theory were not really necessary or, at least, could
be greatly weakened or just avoided.

A typical example is the {\it ergodic hypothesis}. It is interesting
for us a very short history of it (as I see it). Early in his works
Boltzmann started publishing the first versions of the {\it heat
theorem}. The theorem says that one can define in terms of time
averages of total or kinetic energy, of density, and of average
momentum transfer to the container walls, quantities that one could
call {\it specific internal energy} $u$, {\it temperature} $T$, {\it
specific volume} $v$, {\it pressure} $p$ and when two of them varied,
say the specific energy and volume by $du$ and $dv$, they verify:

\begin{equation}\frac{du+pdv}T={\rm\ exact\ }\label{1.1}\end{equation}

At the beginning this was discussed in very special cases (like free
gases). But about fifteen years later Helmoltz noted in a series of
four ponderous papers that for a class of very special systems, that
he called {\it monocyclic}, in which all motions were periodic
and in a sense nondegenerate, one could give appropriate names,
familiar in macroscopic thermodynamics, to various mechanical averages
and then check that they verified the relations that would be expected
between thermodynamic quantities with the same name.

Helmoltz' assumptions about monocyclicity are very strong and I do not
see them satisfied other than in confined one dimensional Hamiltonian
systems. Here is an example of Helmotlz' reasoning (as reported by
Boltzmann).

Consider a $1$--dimensional system with potential $\varphi(x)$ such that
$|\varphi'(x)|>0$ for $|x|>0$, $\varphi''(0)>0$ and $\varphi(x)\,\,
{\vtop{\ialign{#\crcr\rightarrowfill\crcr
\noalign{\kern0pt\nointerlineskip}\hglue3.pt${\scriptstyle
{x\to\infty}}$\hglue3.pt\crcr}}}\,\,+\infty$
(in other words a $1$--dimensional system in a confining
potential). There is only one motion per energy value (up to a shift
of the initial datum along its trajectory) and all motions are
periodic so that the system is {\it monocyclic}. We suppose that the
potential $\varphi(x)$ depends on a parameter $V$.

One defines {\it state} a motion with given energy $E$ and given
$V$. And:
\*

\halign{#\ $=$\ & #\hfill\cr
$U$ & total energy of the system $\equiv  K+\varphi$\cr
$T$ & time average of the kinetic energy $K$\cr
$V$ & the parameter on which $\varphi$ is suposed to depend\cr
$p$ & $-$ time average of $\partial_V \varphi$\cr}
\*
\noindent{}A state is parameterized by $U,V$ and if such parameters 
change by
$dU, dV$ respectively we define:

\begin{equation}dL=-p dV,\qquad dQ=dU+p dV\label{1.1a}\end{equation}
then:
\vskip3pt
\noindent{}{\sl Theorem} (Helmoltz): {\it the differential 
$({dU+pdV})/{T}$ is exact.}
\vskip3pt
In fact let:

\begin{eqnarray}S=&2\log \int_{x_-(U,V)}^{x_+(U,V)}
\sqrt{K(x;U,V)}dx=\nonumber\\
=&2\log \int_{x_-(U,V)}^{x_+(U,V)}
\sqrt{U-\varphi(x)}dx\label{1.2}\end{eqnarray}
($\frac12S$ is the logarithm of the action), so that:

\begin{equation}dS=\frac{\int (dU-\partial_V\varphi(x) dV)
\frac{dx}{\sqrt{K}}}{
\int K\frac{dx}{\sqrt{K}}}\label{1.3}\end{equation}
and, noting that $\frac{dx}{\sqrt K} =\sqrt{\frac2m} dt$, we see that the
time averages are given by integrating with respect to $\frac{dx}{\sqrt
K}$ and dividing by the integral of $\frac{1}{\sqrt K}$. We find
therefore:

\begin{equation}dS=\frac{dU+p dV}{T}\label{1.4}\end{equation}

Boltzmann saw that this was not a simple coincidence: his interesting
(and healthy) view of the continuum (which he probably never really
considered more than a convenient artifact, useful for computing
quantities describing a discrete world) led him to think that in some
sense {\it monocyclicity was not a strong assumption}.

Motions tend to recurr (and they do in systems with a discrete phase
space) and in this light monocyclicity would simply mean that, waiting
long enough, the system would come back to its initial state. Thus its
motion would be monocyclic and one could try to apply Helmoltz' ideas
(in turn based on his own previous work) and perhaps deduce the heat
theorem in great generality.  The nondegeneracy of monocyclic systems
becomes the condition that for each energy there is just one cycle and
{\it the motion visits successively all} (discrete) {\it phase space
points}.

Taking this viewpoint one had the possibility of checking that in all
mechanical systems one could define quantities that one could name
with ``thermodynamic names'' and that would verify properties
coinciding with those that Thermodynamics would predict for
quantities with the same name.

He then considered the two body problem, showing that the
thermodynamic analogies of Helmoltz could be extended to systems which
were degenerate, but still with all motions periodic. This led to
somewhat obscure considerations that seemed to play an important role
for him, given the importance he gave them. They certainly do not help
in encouraging reading his work: the breakthrough paper of 1884
\cite{[B84]}, starts with associating quantities with a thermodynamic
name to Saturn rings (regarded as rigid rotating rings!) and
checking that they verify the right relations (like the second
principle, see Eq.(\ref{1.1})).

In general one can call {\it monocyclic} a system with the property
that there is a curve $\ell\to x(\ell)$, parameterized by its
curvilinear abscissa $\ell$, varying in an interval $0< \ell< L(E)$,
closed and such that $ x(\ell)$ covers all the positions compatible
with the given energy $E$.

Let $ x= x(\ell)$ be the parametric equations so that the energy
conservation can be written:

\begin{equation}\frac12 m\dot\ell^2+\varphi(x(\ell))=E
\label{1.6}\end{equation}
then if we suppose that the potential energy $\varphi$ depends on a
parameter $V$ and if $T$ is the average kinetic energy,
$p=-\langle{\partial_V \varphi}\rangle$ it is, for some $S$:
\begin{equation}dS=\frac{dE+pdV}{T},\qquad p=-\langle{\partial_V
\varphi}\rangle,\quad T=\langle{K}\rangle\label{1.7}\end{equation}
where $\langle \cdot\rangle$ denotes time average.

A typical case to which the above can be applied is the case in which
the whole space of configurations is covered by the projection of a
single periodic motion and the whole energy surface consists of just
one periodic orbit, or at least only the phase space points that are
on such orbit are observable. Such systems provide, therefore, natural
models of thermodynamic behaviour.

Noting that a chaotic system like a gas in a container of volume $V$,
which can be regarded as a parameter on which the potential $\varphi$
(which {\it includes} interaction with the container walls) depends,
will verify ``for practical purposes'' the above property, we see that
we should be able to find a quantity $p$ such that $dE+p dV$ admits
the average kinetic energy as an integrating factor.

On the other hand the distribution generated on the surface of
constant energy by the time averages over the trajectory should be an
invariant distribution and therefore a natural candidate for it is the
uniform distribution, {\it Liouville's distribution}, on the surface
of constant energy. The only one if we accept the viewpoint, problably
Boltzmann's, that phase space is discrete and motion on the energy
surface is a monocyclic permutation of its finitely many cells
(ergodic hypothesis).  It follows that if $\mu$ is the Liouville
distribution and $T$ is the average kinetic energy with respect to
$\mu$ then there should exist a function $p$ such that $T^{-1}$ is the
integrating factor of $dE+p dV$.

Boltzmann shows that this is the case and, in fact, $p$ is the average
momentum transfer to the walls per unit time and unit surface, {\it
ie} it is the {\it physical} pressure.

Clearly this is not a proof that the equilibria are described by the
microcanonical ensemble. However it shows that for most systems,
independently of the number of degrees of freedom, one can define a
{\it mechanical model of thermodynamics}.  

Thermodynamic relations are {\it very general} and simple consequences
of the structure of the equations of motion. They hold for small and
large systems, from $1$ degree of freedom to $10^{23}$ degrees. The
above arguments, based on a discrete view of phase space, suggest that
they hold in some approximate sense (as we have no idea on the precise
nature of the discrete phase space). But they may hold {\it exactly}
even for small systems, if suitably formulated: for instance in the
1884 paper\cite{[B84]} Boltzmann shows that in the {\it canonical
ensemble} the relation Eq.(\ref{1.1}) ({\it ie} the second principle)
holds {\it without corrections} even if the system is small.

Thus the ergodic hypothesis does help in finding out why there are
mechanical ``models'' of thermodynamics: they are ubiquitous in small
and large systems. But such relations are of interest in large systems
and not really in small ones.

For large systems any theory claiming to rest on the ergodic
hypothesis may seem bound to fail because if it is true that a system is
ergodic, it is also true that the time the system takes to go through
one of its cycles is simply too long to be of any interest and
relevance: this was pointed out very clearly by Boltzmann\cite{[B96]}
and earlier by Thomson.

The reason we observe approach to equilibrium over time scales far
shorter than the recurrence times is due to the property that the
microcanonical ensemble is such that {\it on most of phase space the
actual values of the observables, whose averages yield the pressure and
temperature and the few remaining thermodynamic quantities, assume the
same value}\cite{[L]}. This implies that such value coincides with the
average and therefore verifies the {\it heat theorem} if $p$ is the
pressure (defined as the average momentum transfer to the walls per
unit time and unit surface).

The ergodic hypothesis loses it importance and fundamental nature and
it appears simply as a tool used in understanding that some of the
relations that we call ``macroscopic laws'' hold in some form for {\it
all} systems, whether small or large.
\end{section}

\begin{section}{The Chaotic hypothesis}

A natural question is whether something similar to the above
development can be achieved in systems out of equilibrium. Here I am
not thinking of systems evolving in time: rather I refer to properties
of systems that reach a stationary state under the influence of
external non conservative forces acting on them. For instance I think
of an electric circuit in which a current flows (stationarily) under
the influence of an electromotive field. Or of a metal bar with two
different temperatures fixed at the extremes. Or of a Navier--Stokes
fluid in a Couette flow.

The first two systems, regarded as microscopic systems ({\it ie} as
mechanical systems of particles), do certainly have a very chaotic
microscopic motions even in absence of external driving (while
macroscopically they are in a stationary state and nothing happens,
besides a continuous, sometimes desired, heat transfer from the system
to the surroundings). The third system also behaves, as a macroscopic
system, very chaotically at least when the Reynolds number is large.

Can one do something similar to what Boltzmann did?

The first problem is that the situation is quite different: there is
no established nonequilibrium Thermodynamics to guide us. The great
progresses of the theory of stationary nonequilibrium that took place
in the past century (I mean the XX), at least the ones that are
unanimously recognized as such, only concern properties of {\it
incipient} non equlibrium: {\it ie} transport properties at vanishing
external fields (I think here of Onsager's reciprocity and its
quantitative form given by Green--Kubo's transport theory). So it is
by no means clear that there is any general non equilibrium
thermodynamics.

Nevertheless in 1973 a first suggestion that a general theory might be
possible for non equilibrium systems in stationary and chaotic states
was made by Ruelle in talks and eventually written down in
papers\cite{[R1]}.

The proposal is very ambitious as it suggests a {\it general and
essentially unrestricted} answer to which should be the ensemble that
describes stationary states of a system, {\it whether in
equilibrium or not}.

The ergodic hypothesis led Boltzmann to the general theory of
ensembles (as acknowledged by Gibbs, whose work has been perhaps the
main channel through which the allegedly obscure works of Boltzmann
reached us): besides giving the second law, Eq.(\ref{1.1}), it also
prescribed the microcanonical ensemble for describing equilibrium
statistics.

The reasoning of Ruelle was that from the theory of simple chaotic
systems one knew that such systems, for the simple fact that they are
chaotic, will reach a ``unique'' stationary state. Therefore simply
assuming chaoticity would be tantamount to assuming that there is a
uniquely defined ensemble which should be used to compute the
statistical properties of a system out of equilibrium.

Therefore one is, in a very theoretical way, in a position to inquire
whether such unique ensemble has {\it universal properties} valid for
small and large systems alike: of course we cannot expect too many of
them to hold. In fact in equilibrium theory the only one I know is
precisely the heat theorem, besides a few general (related)
inqualities ({\it eg positivity of the specific heat or of
compressibility}). The theorem leads, indirectly as we have seen, to
the microcanonical ensemble and then, after one century of work, to a
rather satisfactory theory of phenomena like phase transitions, phase
coexistence, universality.

In the end the role of the ergodic hypothesis emerges, at least in my
view, as greatly enhanced: and the idea of Ruelle seems to be its
natural (and I feel unique) extension out of equilibrium.

Of course this would suffer from the same objections that are
continuously raised about the ergodic hypothesis: namely ``there is
the time scale problem''.

To such objections I do not see why the answer given by Boltzmann
should not apply {\it unchanged}: large systems have the extra
property that the interesting observables take the same value in the
whole (or virtually whole) phase space. {\it Therefore they verify any
relation that is true no matter whether the system is large or small}:
such relations (whose very existence is, in fact, surprising)
might be of no interest whatsoever in small systems (like in the above
mentioned Boltzmann's rigid Saturn ring, or in his other similar
example of the Moon regarded as a rigid ring rotating about the
Earth).

Ruelle's proposal was formulated in the case of fluid mechanics: but
it is so clearly more general that the reason why it was not
explicitly proposed for statistical systems is probably due to the
fact that, as a principle, it required some ``check'' if formulated
for Statistical MEchanics: as originally stated and without any
further check it would have been analogous, in my view, to the ergodic
hypothesis without the heat theorem (or other consequences drawn from
the theory of statistical ensembles). 

Evidence for the non trivial applicability of the hypothesis built up
quite rapidly and it was repeatedly hinted in various papers dealing
with numerical experiments, mostly on very small particle systems
($<100$ to give an indication)\cite{[H]}. In attempting at
underdstanding one such experiment\cite{[ECM2]} the following
``formal'' interpretation of the Ruelle's priciple was
formulated\cite{[GC]} for statistical mechanics (as well as for fluid
mechanics, replacing ``many particles system'' with ``turbulent
fluid'')) in the form: \vskip3pt

{\it Chaotic hypothesis: A many particle system in a stationary state
can be regarded as a transitive Anosov system {\it(see below)} for the
purpose of computing the macroscopic properties of the system.}
\vskip3pt

The hypothesis was made first in the context of reversible systems
(which were the subject of the experimental work that we were
attempting to explain theoretically). The assumption that the
system is Anosov (see below) has to be interpreted when the system
has an attractor strictly smaller than the available phase space ({\it ie}
not dense in it), as saying that the attractor itself can be
regarded as a smooth Anosov systems (see below).

The latter interpretation {\it rules out} fractal attractors and, to
include them, it could be replaced by changing ``Anosov'' into ``Axiom
A'': but I prefer to wait if there is real need of such an
extension. It is certainly an essential extension for small systems,
but it is not clear to me how relevant could fractality be when the
system has $10^{23}$ particles).

A transitive Anosov system is a {\it smooth} system with a dense orbit
(the latter condition is to exclude trivial cases, like when the
system consists of two chaotic but noninteracting subsystems) and such
that around every point $x$ one can set up a local coordinate system
that a) {\it depends continuously on $x$ and is covariant} ({\it ie}
it follows $x$ in its evolution) and b) is {\it hyperbolic} ({\it ie}
transversally to the phase space velocity of any chosen point $x$ the
motion of nearby points looks, when seen from the coordinate frame
covariant with $x$, as a hyperbolic motion near a fixed point.

This means that on (each) plane transversal to the phase space velocity
of $x$ there will be a ``stable coordinate surface'', the {\it stable
manifold} through $x$, whose points trajectories get close to the
trajectory of $x$ at exponential speed as the time tends to $+\infty$
and an ``unstable coordinate surface'', the {\it unstable manifold},
whose trajectories get close to the trajectory of $x$ at exponential
speed as the time tends to $-\infty$. The direction parallel to the
velocity can be regarded as a {\it neutral} direction where, in the
average, no expansion or contraction occurs.

Anosov systems are the {\it paradigm} of chaotic systems: they are the
analogues of the harmonic oscillators for ordered motions. Their
simple but surprising and deep properties are by and large very well
understood; particularly in the discrete time cases that we consider
below. Unfortunately they are not as well known as they should
among physicists, who seem confused by the language in which they are
usually presented: however it is a fact that such a remarkable
mathematical object has been introduced by mathematicians and the
physicists must therefore make an effort at understanding the new
notion and its physical significance.

In particular, if a system is Anosov: {\it for all} observables $F$
({\it ie} continuous functions on phase space) and for almost all
initial data $x$ the time average of $F$ exists and can be computed by
a phase space integral with respect to a distribution $\mu$ uniquely
determined on phase space ${\cal F}$:

\begin{equation}\lim_{T\to\infty} 
\frac1T \int_0^T F(S_t x) \,dt=\int_{\cal F}
F(y)\,\mu(dy)\label{2.1}\end{equation}
{\it``almost all''} means apart from a set of zero volume in phase
space.  The distribution is called the SRB distribution: it was proven
to exist by Sinai\cite{[SRB]} for Anosov systems and the result was
extended to the much more general Axiom A attractors by Ruelle and
Bowen\cite{[R1]}. {\it Natural distributions} were, independently,
discussed and shown to exist\cite{[LY]} for other (related and simpler)
dynamical systems.

Clearly the chaotic hypothesis solves {\it in general} ({\it ie} for
systems that can be regarded as ``chaotic'') the problem of
determining which is the ensemble to use to study the statistics of
stationary systems in or out of equilibrium (it clearly implies the
ergodic hypothesis in equilibrium), in the same sense in which the
ergodic hypothesis solves the equilibrium case. 

Therefore the first problem with such an hypothesis is that it will be
very hard to prove it in a mathematical sense: the same can be said
about the ergodic hypothesis which is not only unproved for most
cases, but it will remain such, in systems of statistical mechanical
interest, for long if not forever, aside from some very special cases
(like the hard core gas). The chaotic hypothesis might turn out to be
false in interesting cases, like the ergodic hypothesis which does not
hold for the simplest systems studied in statistical mechanics, like
the free gas, the harmonic chain and the black body radiation. Worse:
it is {\it known} to be false for trivial reasons in some systems in
equilibrium (like the hard core gas): simply because the Anosov
definition requires smoothness of the evolution and systems with
collisions are not smooth systems (in the sense that the trajectories
are not differentiable as functions of the initial data).

However, interestingly enough, the case of hard core systems is
perhaps the system closest to an Anosov system that can be thought of
and that is also of statistical mechanical relevance. To an extent
that there seem to be no known properties that such system does not
share with an Anosov system. Aside from the trivial fact that it is
not a smooth system, the hard core system behaves, for Statistical
Mechanics purposes, {\it as if it was a Anosov system}. Hence it is
the prototype system to study in looking for applications of the
chaotic hypothesis.

The {\it problem} that remains is whether the chaotic hypothesis
has any power to tell us something about nonequilibrium statistical
mechanics. This is the real, deep, question for anyone who is willing
to consider it. {\it One consequence} is the ergodic
hypothesis, hence the heat theorem: but this is {\it too little} even
though it is a very important property for a theory with the ambition
of being a {\it general} extension of the theory of equilibrium
ensembles.

I conclude this section with a comment useful in the following. As is
well known by who has ever attempted a numerical (or real) experiment,
one often does not observe systems in continuous time: but rather one
records the state of the system at times when some event that is
considered intereresting or characteristic happens. Calling such
events ``{\it timing events}'' the system then appears as having a
phase space of dimension one unit lower: because the set of timing
events has to be thought of as a surface in phase space transversal to
the phase space velocity of the trajectories $t\to S_t x$. 

If $x$ is a timing event and $\vartheta(x)$ is the time that one has
to wait until the next timing event happens, the time evolution becomes
a map $x\to Sx\equiv S_{\vartheta(x)}x$ of $x$ into the following
timing event. For instance one could record the configuration
of a system of hard balls every time that a collision takes place, and
$S$ will map a collision configuration into the next one.

The chaotic hypothesis can be formulated for such ``Poincar\`e's
sections'' of the continuous time flow in exaclty the same way: and
this is in fact a simpler notion as there will be no ``{\it neutral
direction}'' and the covariant local system of coordinates will be
simply based on a stable and an unstable manifold through every point
$x$.

{\it In the following section we take the point of view that time
evolution has been discretized in the above sense} ({\it ie} via a
Poincar\`e's section on a surface of timing events): this simplifies a
discussion, but in a minor way\cite{[Ge]}.
\end{section}

\begin{section}{Fluctuation theorem for reversibly dissipating systems.}

The key to find applications is that the apparently inconsequential
hypothesis that the system is Anosov provides us not only with an
existence theorem of the SRB distribution $\mu$ but {\it also} with an
explicit expression for it. How explicit? as we shall see not too far
from what we are used to in equilibrium statistical mechanics ({\it
eg} $\mu=e^{-\beta H}$): where apparently unmanageable expressions and
hopeless integrals have important and beautiful applications in spite
of their obvious non computability.

The expression is the following: there is a partition of phase space
into cells $E_1,E_2,\ldots$ which in a sense that I do not specify
here\cite{[G1]} is ``{\it covariant}'' with respect to time evolution
and to the other symmetries of the system (if any: think of parity or
time reversal) such that the average value of an observable can be
computed as:

\begin{equation}\langle{F}\rangle=\int_{\cal F} F(y)\,\mu(dy)
=\frac{\sum_{E_i}\Lambda_{u,T}^{-1}(x_i) F(x_i)} {\sum_{E_i}
\Lambda_{u,T}^{-1}(x_i)}\label{3.1}\end{equation}
where $x_i\in E_i$ is a point suitably chosen in $E_i$ (quite, but not
completely, arbitrarily for technical, trivial,
reasons\cite{[GC],[G1]}) and $\Lambda_{u,T}(x)$ is the expansion of a
surface element lying on the unstable manifold of $S_{-\frac12T}
x$ and mapped by $S_T$ into a surface element around $S_{\frac12T}x$.

Of course Eq.(\ref{3.1}) requires that the cells be so small that $F$
has neglegible variations inside them: if this is not the case then
one simply has to {\it refine} the partiction into smaller cells, until
they become so small that $F$ is a constant inside them (for practical
purposes).  This can be done simply by applying the time evolution map
and its inverse to the partition that we already imagine to have, but
which has large cells, and then intersecting the elements of the new
partitions obtained to get a finer partition. The hyperbolicity of the
evolution implies that the partition into cells can be made as fine as
desired.

Another reason why we need small cells is to insure that the weights
themselves do not depend too much on which point $x_i$ is chosen to
evaluate them: the precise condition is somewhat
delicate\cite{[boundary]}.

An example of an application of the above formula is obtained by
studying the phase space volume contraction rate $\sigma(x)$: this is
defined as the logarithm of the Jacobian determinant $\Lambda(x)$ of
the time evolution map (recall that we are now considering a discrete
time evolution $S$, as explained at the end of the preceding
section). Suppose that we ask for the fluctuations of the average of
the ``dimensionless contraction'' $\sigma(x)/\sigma_+$ where
$\sigma_+$ is the (infinite) time average $\sigma_+=\int
\sigma(y)\,\mu(dy)$, that is assumed strictly positive (it could be
zero, for instance in a equilibrium system where the evolution is
Hamiltonian and conserves volume in phase space; {\it but} it
cannot\cite{[R2]} be $<0$). The positivity of the time average of
$\sigma$ can be taken as the very definition of ``dissipative''
motions.

This is the quantity: $p=\frac1{\tau\sigma_+}\sum_{k=-
\frac12\tau}^{\frac12\tau} \sigma(S^k x)$. It will have a probability
distribution, in the stationary state, that we write $\pi_\tau(p)$. We
now compare $\pi_\tau(p)$ to $\pi_\tau(-p)$, which is clearly a ratio
of probabilities of two events one of which will have an extremely
small probability (the expected value of $p$ being $1$).

{\it Suppose that the system is time reversible}: {\it ie} that there is
an isometry of phase space $I$ that anticommutes with the evolution:
$IS=S^{-1}I$. Then:

\begin{equation}\frac{\pi_\tau(p)}{\pi_\tau(-p)}=\frac {\sum_{E_i;
p}\Lambda_{u,T}^{-1}(x_i)} {\sum_{E_i;
-p}\Lambda_{u,T}^{-1}(x_i)}\label{3.2}\end{equation}
where the sum in the numerator extends over the cells $E_i$ in which
the total dimensionless volume contraction rate is $p$ anf the sum in
the denominator over those with contraction rate $-p$.

Here we take $T=\tau$ for the purpose of a partial illustration: this
is {\it not} allowed and in a sense it is the {\it only} difficulty
in the discussion. But taking $T=\tau$ conveys some of the main
ideas. If this ``interchange of limits'' is done then one simply
remarks that the sum in the denominator of Eq.(\ref{3.2}) can be
performed over the same cells as that in the numerator, provided we
evaluate the weight in the denominator at the point $I x_i$, {\it ie}
provided we use in the denominator the weight
$\Lambda_{u,T}^{-1}(Ix_i)$: this is so because time reversal maps a
cell in which the dimensionless rate of volume contraction is $p$ into
one in which it is $-p$ and viceversa. But time reversal also
interchanges expasion and contraction so that
$\Lambda_{u,T}^{-1}(Ix_i)=\Lambda_{s,T}(x_i)$, if the contraction rate
along the stable manifold $\Lambda_{s,T}$ is defined in the same way
as $\Lambda_{u,T}$ by exchanging stable and unstable manifolds. This
means that the ratio between corresponding terms is now
$\Lambda_{u,T}^{-1}(x_i)\Lambda_{s,T}^{-1}(x_i)$.

Since the latter quantity is essentially the {\it total contraction
rate} up to a factor bounded independently of the value of $T$
(because the angle between the stable and unstable manifolds is
bounded away from zero by the continuity property of Anosov systems)
it follows that the ratio Eq.(\ref{3.2}), in this (rather
uncontrolled) approximation $T=\tau$, is $\tau p \sigma_+$ {\it ie}
simply the contraction rate which has the {\it same value for all
cells considered}, by construction.  Conclusion:

\begin{equation}\lim_{\tau\to\infty}\frac1{\tau\,p\,\sigma_+}
\log\frac{\pi_\tau(p)}{\pi_\tau(-p)}=1
\label{3.3}\end{equation}
which is the {\it fluctuation theorem} if $\pi_\tau$ are evaluated with
respect to the SRB distribution of the system.

The above ``proof'' is missing a key point: namely the interchange of
limits. Fixing $\tau=T$ means that we are not computing the
probabilities in the SRB distribution but, at best, in some
approximation of it. In experimental tests one need the theorem to
hold when the limits are taken in the proper order ({\it ie} first
$T\to\infty$ and {\it then} $\tau\to\infty$). The latter theoretical
aspects have been discussed in the original papers\cite{[GC]}, where
it is shown that the limit is approached as $\tau^{-1}$; and formal
proofs are also available\cite{[G2],[R3]}.

That this is not a fine point of rigor can be seen from the fact that
if one disregards it then other proofs of the ``same'' result {\it
but} with $\tau=T$ become possible. In other words the result has a
``tendency'' to be general\cite{[MR],[K]} but it can be proved in the
right form of Eq(\ref{3.3}) only under strong chaoticity
assumptions. It is very interesting that in weaker forms a result
closely related to the fluctuation theorem can be obtained for {\it
completely different} dynamical systems {\it ie} for stochastic
evolutions\cite{[K]}. It is possible that for the stochastic evolution
the result could be extended to become a closer analogue of the above,
solving the mentioned problem of the interchange of limits: one would,
in fact, think that the noise makes the system as chaotic as one may
possibly hope.

The result Eq.(\ref{3.3}), has to be tested because in all applications
we do not know whether the system is Anosov and to what extent it can
be assumed such. And its verification provides a form of test of the
chaotic hypothesis.

Other equivalent formulations of the fluctuation theorem are in terms
of the ``free energy'' of the observable $p$:
$\zeta(p)=\lim_{\tau\to\infty} \frac1\tau\log \pi_\tau(p)$; it
becomes:

\begin{equation}\frac{\zeta(p)-\zeta(-p)}{p\sigma_+}=1
\label{3.4}\end{equation}
which says that the odd part of $\zeta(p)$ is linear in $p$ with a
{\it determined and parameter free}, slope: note that without
reversibility one could only expect that $\zeta(p)$ had a quadratic
maximum at $p=1$ ({\it central limit theorem} for the observable
$\sigma(x)$) which stays quadratic as long as
$|p-1|=O(\frac1{\sqrt{{\tau}}})$. The fluctuation theorem instead
gives informations concerning huge deviations $|p-1|=O(2)$! it is a
{\it large deviation theorem}.

The main interest, so far, of the above theorem is that it has shown
that Ruelle's principle has some power of prediction. In fact the
result has been checked in various small systems\cite{[BGG]}. The
first of which was its experimental discovery\cite{[ECM2]} {\it
preceding} the chaotic hypothesis and fluctuation theorem
formulations.

It is also interesting because of its {\it universal validity}: it is
system independent (provided reversible), hence it is a general law
that should be satisfied if the chaotic hypotesis is the correct
mathematical translation of our intuitive notion of chaos, and Anosov
systems catch it fully.

The question whether the above results can also be obtained from the
chaotic hypothesis formulated in terms of the continuous time flow on
phase space (rather than for a map between timing events, see the last
comments in the previous section) would leave us unhappy if it did not
have a positive answer: it does have a positive answer\cite{[Ge]}.
\end{section}

\begin{section}{Onsager's reciprocity and Green--Kubo's formula.}

The fluctuation theorem degenerates in the limit in which $\sigma_+$
tends to zero, {\it ie} when the external forces vanish and
dissipation disappears (and the stationary state becomes the
equilibrium state).

Since the theorem deals with systems that are time reversible {\it at
and outside} equilibrium Onsager's hypotheses are certainly verified
and the system should obey reciprocal response relations at vanishing
forcing. This led to the idea that there might be a connection between
fluctuation theorem and Onsager's reciprocity and also to the related
(stronger) Green--Kubo's formula.

This is in fact true: if we define the {\it microscopic thermodynamic
flux} $j(x)$ associated with the {\it thermodynamic force} $E$ that
generates it, {\it ie} the parameter that measures the strength of the
forcing (which makes the system not Hamiltonian), via the relation:

\begin{equation}
j(x)=\frac{\partial\sigma(x)}{\partial E}\label{4.1}
\end{equation}
(not necessarily at $E=0$) then in [G2] a heuristic proof shows that
the limit as $E\to0$ of the fluctuation theorem becomes simply (in the
continuous time case) a property of the average, or ``macroscopic'',
{\it flux} $J=\langle{j}\rangle_{\mu_E}$:

\begin{equation}\frac{\partial J}{\partial E}\big|_{E=0}=\frac12
\int_{-\infty}^{\infty} \langle{j(S_tx)j(x)}\rangle_{\mu_E}\Big|_{E=0}
\,dt\label{4.2}\end{equation}
where $\langle{\cdot}\rangle_{\mu_E}$ denotes average in the stationary
state $\mu_E$ ({\it ie} the SRB distribution which, at $E=0$, is simply
the microcanonical ensemble).

If there are several fields $E_1,E_2,\ldots$ acting on the system we
can define several thermodynamic fluxes $j_k(x){\buildrel def\over =}
\partial_{E_k}\sigma(x)$ and their averages $\langle{j_k}\rangle_\mu$:
a simple extension of the fluctuation theorem\cite{[G3]} is shown to
reduce, in the limit in which all forces $E_k$ vanish, to:

\begin{eqnarray}L_{hk}{\,{\buildrel def\over=}\,}
\frac{\partial J_{h}}{\partial
E_k}\big|_{E=0}=\kern1cm{}\nonumber\end{eqnarray}
\begin{equation}=\frac12
\int_{-\infty}^{\infty} \langle{j_h(S_tx)j_k(x)}\rangle_{E=0}
\,dt=L_{kh}\label{4.3}\end{equation}
therefore we see that the fluctuation theorem can be regarded as {\it
an extension to non zero forcing} of Onsager's reciprocity and,
actually, of Green--Kubo's formula.

Certainly assuming reversibility in a system out of equilibrium can be
disturbing: therefore one can inquire if there is a more general
connection between the chaotic hypothesis and Onsager's reciprocity and
Green--Kubo's formula. This is indeed the case and provides us with a
{\it second application} of the chaotic hypothesis valid, however, only in
zero field. It can be shown that the relations Eq.(\ref{4.3}) follow
from the sole assumption that at $E=0$ the system is time reversible and
that it verifies the chaotic hypothesis at $E=0$: at $E\ne0$ it can
be, as in Onsager's theory, not reversible\cite{[GR]}.

It is not difficult to see, technically, how the fluctuation theorem,
in the limit in which the driving forces tend to $0$, formally yields
Green--Kubo's formula.

We consider time evolution in continuous time and simply note that
Eq.(\ref{3.3}) implies that, for all $E$ (for which the system is
chaotic):

\begin{equation}
\lim_{\tau\to\+\infty} \frac1\tau
\log \langle e^{I_E}\rangle_{\mu_E}=0\label{4.4}
\end{equation}
where $I_E\,{\buildrel def \over =}\,\int \sigma(S_tx) dt$ with
$\sigma(x)$ being the divergence of the equations of motion ({\it ie}
the phase space contraction rate, in the case of continuous
time). This remark\cite{[Bo]} (that says that essentialy $\langle
e^{I_E}\rangle_{\mu_E}\equiv1$ or more precisely it is not too far
from $1$ so that Eq.(\ref{4.4}) holds) can
be used to simplify the analysis in {}\cite{[G3]} as follows.

Differentiating both sides with respect to $E$, not worrying about
interchanging derivatives and limits and the like, one finds that the
second derivative with respect to $E$ is a sum of six terms. Supposing
that for $E=0$ the system is Hamiltonian and (hence) $I_0\equiv 0$,
the six terms are, when evaluated at $E=0$:

\begin{eqnarray}
&\frac1\tau\langle \partial^2_E I_E\rangle_{\mu_E}|_{E=0}
-\frac1\tau\langle (\partial_E
I_E)^2\rangle_{\mu_E}|_{E=0}+\nonumber\\ &+\frac1\tau\int \partial_E
I_E(x) \partial\mu_E(x)|_{E=0}+\nonumber\\ &-\frac1\tau\Big(\langle
(\partial_E I_E)^2\rangle_{\mu_E}\cdot \int 1\,\partial_E
\mu_E\Big)|_{E=0}+\label{4.5}\\ &\frac1\tau\int \partial_E I_E(x)
\partial\mu_E(x)|_{E=0}
+\frac1\tau\int 1\cdot\partial_E^2\mu_E|_{E=0}\nonumber
\end{eqnarray}
and we see that the fourth and sixth terms vanish being derivatives of
$\int \mu_E(dx)\equiv 1$, the first vanishes (by integration by parts)
because $I_E$ is a divergence and $\mu_0$ is the Liouville
distribution (by the assumption that the system is Hamiltonian at
$E=0$ and chaotic). Hence we are left with:

\begin{equation}
\Big(-\frac1\tau\langle (\partial_E I_E)^2\rangle_{\mu_E}+\frac2\tau
\int \partial_E I_E(x) \partial_E
\mu_E(x)\Big)_{E=0}=0\label{(4.6)}\end{equation}
where the second term is, since the distribution $\mu_E$ is
stationary, $2\tau^{-1} \partial_E ( \langle \partial_E
I_E\rangle_{\mu_E})|_{E=0}\equiv 2\partial_E J_E|_{E=0}$; and the
first term tends to $\int_{-\infty}^{+\infty} \langle j(S_t x)
j(x)\rangle_{E=0} dt$ as $\tau\to\infty$. Hence we get Green--Kubo's
formula in the case of only one forcing paprameter.

The argument should be extended to the case in which $E$ is a vector
describing the strength of various driving forces acting on the
system\cite{[G3]}: but one needs a generalization of Eq({4.4}). The
latter is a consequence of the fluctuation theorem, but the theorem
had to be extended\cite{[G3]} to derive also Green--Kubo's formula
(hence reciprocity) when there were several independent forces acting
on the system..

The above analysis is unsatisfactory because we interchange limits and
derivatives quite freely and we even take derivatives ot $\mu_E$,
which seems to require some imagination as $\mu_E$ is concentrated on
a set of zero volume. On the other hand, under the strong hypotheses
which we suppose to be (that the system is Anosov), we should not need
extra assumptions. Indeed the above mentioned non heuristic
analysis\cite{[GR]} is based on the study on the differentiability of SRB
distributions with respect to parameters\cite{[Rdiff]}.

A {\it third application} of the chaotic hypothesis, still limited to
reversible systems, is the following: consider the probability that
certain observables $O_1,O_2,\ldots$ are measured during a time
interval $[-\frac11\tau,\frac12\tau]$ during which the system evolves
between the point $S_{-\frac12\tau}x$ and $S_{\frac12\tau}x$. And
suppose that we see the {\it path} or {\it pattern} $\omega$ given by
$t\to O_1(S_tx),O_2(S_tx),\ldots$.

Assuming, for simplicity, that $O_j$ are {\it even} under time
reversal the ``time reversed'' pattern $I\omega$ will be $t\to
O_1(S_{-t}x), O_2(S_{-t}x)$ and it will be clearly {\it very
unlikely}. Suppose that we look at the relative probabilities of
various patterns {\it conditioned} to an average (over the time
interval $[-\frac11\tau,\frac12\tau]$) dimensionless volume
contraction rate $p$. Then one can prove\cite{[G4]}, under the chaotic
hypothesis, that the relative probabilities of patterns in presence of
rate $p$ is {\it the same as that of the time reversed patterns in
presence of rate} $-p$.

Since the contraction rate of volume in phase space can be interpreted
as {\it entropy creation rate}, as suggested for instance by the above
use, Eq.(\ref{1.3}) of the phase space contraction to define the
thermodynamic fluxes, as ``conjugate'' observables to the external
thermodynamic forces, the latter statement has some interest as it can
be read as saying that ``it costs no extra effort to realize events
normally regarded as impossible once one succeeds in the enterprise of
reversing the sign of entropy creation rate''\cite{[G4]}.

The interpretation of phase space contraction rate as {\it entropy
creation rate} meets opposition, fierce at times: however it seems to
me a very reasonable proposal for a concept that we should not forget
has not yet received a universally accepted definition and therefore
its definition should at least be considered as an open problem.
\vskip3pt

\end{section}

\begin{section}{Reversible versus irreversible dissipation.
Nonequilibrium ensembles?}

A system driven out of equilibrium can reach a stationary state
(and not steam out of sight) only if enough dissipation is
present. This means that any mechanical model of a system reaching a
stationary state out of equilibrium {\it must} be a model with non
conservative equations of motion in which forces representing the
action of the thermostats, that keep the system from heating up, are
present.

Thus a generic model of a system stationarily driven out of
equilibrium will be obtained by adding to Hamilton's
equations (corresponding to the non driven system) other terms 
representing forces due to the thermostat action.

Here one should avoid attributing a fundamental role to special
assumptions about such forces. One has to realize that there is {\it
no privileged} thermostat. One can consider many of them and they
simply describe various ways to take out energy from the system.

Thus one can use stochastic thermostats, and there are many types
considered in the literature; or one can consider deterministic
thermostats and, among them, reversible ones or irreversible ones.

Each thermostat requires its own theory. However the same system may
behave in the same way under the action of different thermostatting
mechanisms: if the only action we make on a gas tube is to keep the
extremes temperatures fixed by taking in or out heat from them the
difference may be irrelevant, at least in the limit in which the tube
becomes long enough and as far as what happens in the middle of it is
concerned.

But of course the form of the stationary state may be very different
in the various cases, even when we think that the differences are only
minor boundary effects. For instance, in the case of the gas tube, if
our model is of deterministic dissipation we expect that the SRB state
be concentrated on a set of {\it zero phase space volume} (because
phase space will in the average contract, when $\sigma_+>0$, so that
any stationary state has to be concentrated on a set of zero volume,
{\it which however could still be dense} and ususally will be). While
if the model is stochastic then the stationary state will be described
by a {\it density} on phase space. Nothing could seem more different.

Nevertheless it might be still true that in the limit of an infinite
tube the two models give the same result: in the same sense as the
canonical and microcanonical ensembles describe the same state even
though the microcanonical ensemble is supported on the energy surface,
which has zero volume if measured by using the canonical ensemble
(which is given by a density over the whole available phase space).

Therefore we see that out of equilibrium we have in fact {\it much more
freedom to define equivalent ensembles}. Not only we have (very
likely) the same freedom that we have in equilibrium (like fixing the
total energy or not, or fixing the number of particles or not, passing
from microcanonical to canonical to grand canonical {\it etc}) but
{\it we can also change the equations of motion and obtain different
stationary states, {\it ie} different SRB distributions, which will
however become the same in the thermodynamic limit}.

Being able to prove the mathematical equivalence of two thermostats
will amount at proving their physical equivalence. This again will be
a difficult task, in any concrete case.

What I find fascinating is that the above remarks provide us with the
possibility that a {\it reversible thermostat can be equivalent in the
thermodynamic limit to an irreversible one}. I conclude by
reformulating a conjecture, that I have already stated many times in
talks and in writings\cite{[G5]}, which clarifies the latter statement.

Consider the following two models describing a system of hard balls
in a periodic (large) box in which there is a lattice of obstacles
that forbid collisionless paths (by their arrangement and size): the
laws of motion will be Newton's laws (elastic collisions with the
obstacles as well as between particles) plus a constant force $E$
along the $x$--axis {\it plus a thermostatting force}.

In the first model the thermostatting force is simply a constant times
the momentum of the particles: it acts on the $i$-tth particle as
$-\nu p_i$ if $\nu$ is a ``friction'' constant. Another model is a
force proportional to the momentum but via a proportionality factor
that is not constant and depends on the system configuration: it has
the form $-\alpha(x)p_i$ with $\alpha(x)=E\cdot\sum_i p_i/\sum_i
p_i^2$.

The first model is essentially the model used by Drude in his theory
of conduction in metals. The second model has been used very often in
recent years for theoretical studies and has thus acquired a respected
status and a special importance: it was among the first models used in
the experiments and theoretical ideas that led to the connection
between Ruelle's ideas for turbulent motion in fluids and
nonequilibrium statistical mechanics\cite{[H]}. I think that the
importance of such works cannot be underestimated: without them the
recent theoretical developments would have been simply unthinkable, in
spite of the fact that {\it a posteriori} they seem quite independent
and one could claim (unreasonably in my view) that everything could
have been done much earlier.

Furthermore the second model can be seen as derived from Gauss' least
constraint principle. It keeps the total (kinetic) energy exactly
constant over time (taking in and out energy, as needed) and is called
{\it Gaussian thermostat}. {\it Unlike the first model it is
reversible}, with time reversal being the usual velocity
inversion. Thus the above theory and results based on the chaotic
hypothesis apply.

My conjecture was (and is) that:
\vskip3pt

1) compute the average energy per particle that the system has in the
constant friction case and call it ${\cal E}(\nu)$ calling also
$\mu_\nu$ the corresponding SRB distribution.

2) call $\tilde \mu_{\cal E}$ the SRB distribution for the Gaussian
thermostat system when the total (kinetic) energy is fixed to the
value ${\cal E}$

3) then $\mu_\nu=\tilde \mu_{{\cal E}(\nu)}$ {\it in the
thermodynamic limit} (in which the box size tends to become infinitely
large but with the number of particles and the total energy
correspondingly growing so that one keeps the density and the energy
density constant) and for {\it local} observables, {\it ie} for
observables that depend only on the particles of the system localized
in a fixed finite region of the container. This means that the
equality takes place in the usual sense of the theory of
ensembles\cite{[R4]}.
\vskip3pt

This opens the way to several speculations as it shows that the
reversibility assumption might be not so strong after all. And
results for reversible systems may carry through to irreversible
ones.

I have attempted to extend the above ideas also to cases of turbulent
motions but I can only give here references\cite{[G5],[G6]}.
\vskip3pt

{\it Acknowledgements: this work has been supported also by Rutgers
University and by the European Network \# ERBCHRXCT940460 on:
``Stability and Universality in Classical Mechanics". I am indebted to
F. Bonetto for several enlightening discussions\cite{[Bo]}.}
\end{section}

\vskip3pt
\noindent{}{\it Internet access: The author's quoted preprints can be
downloaded (latest revision) at:

\centerline{\tt http://chimera.roma1.infn.it}

\noindent{}in the Mathematical Physics Preprints page.

\noindent{}\sl{}Author's e-mail:\ giovanni@ipparco.roma1.infn.it
}

\hbox{}\hfill{\it Typeset using {REV\TeX}}

\begin{references}
\bibitem{[boundary]} The reason can be easily understood by analogy
with a statistical mechanics case. Imagine a $1$--dimensional nearest
neighbour Ising spin system and that $\mu$ is the {\it infinite
volume} zero field Gibbs state with coupling $J$. Then we can use a
formula like Eq.(\ref{3.1}) where $E_i$ are the spin configurations
which have a specific form (one among the $2^T$ possible ones) in the
box $[-\frac12T,\frac12T]$ and are arbitrary outside: clearly to get
$\langle F\rangle|_\mu$ right the function $F$ has to depend only on
few ``centrally located spins'', which means that $F$ has to be
essentially constant in each of the $E_i$'s; hence, as said, the $E_i$
have to be small enough, {\it ie} $T$ large enough. But we also need
that the boundaries have no influence on the expectation values that
we compute. A choice of $x_i\in E_i$ means selecting a spin
configuration inside the box $[-\frac12T,\frac12T]$ {\it and ouside
it}: hence the weight ({\it ie} $e^{-H}$ if $H$ is the energy) depends
on the form of the spin configuration $x_i$ {\it outside} the
box. This means that the relative weights of the possible
configurations inside the box (corresponding to the possible choices
of $x_i$) do depend on $x_i$ and change by order $O(J)$, {\it not
small as $T\to\infty$}, by changing $x_i$ keeping it inside $E_i$ {\it
ie} keeping the same spin configuration in the box.  Therefore care
has to be paid in the choice of $x_i$ so that the relative weights are
``right''. However there are always ``pathological'' choices of $x_i$
for which something ``goes wrong'' for some $F$ unless some care is
exercized in choosing $x_i$. Think of $F$ as being the value of the
spin $\sigma_0$ located at the origin and of choosing $x_i$ so that
the first spins out of the box, $\sigma_{\pm(\frac12T+1)}$, {\it the
same } as the one, $\sigma_0$, at the center of the box: this amounts,
if such a $x_i$ is chosen, at computing the distribution of $\sigma_0$
in a model with an external field $2J$ acting on the spin at site $0$:
this is obviously going to give the wrong expectation value for
$F=\sigma_0$ although in the average we produce the same state as the
correct one. In fact the only wrong averages concern those of spins
located near the origin (note that the analogue of this this would
be false in $2$--dimensions, in the phase transitions region). An
example of a ``correct'' choice of $x_i$, in this analogy, would be
simply to take $x_i$ with
$\sigma_{\pm(\frac12T+1)}=\sigma_{\pm\frac12T}$ and arbitrary
elsehwere outside the box (using that the range is nearest
neighbour). Finally the analogy is not accidental: the theory of the
SRB states is done precisely by showing that they can be thought of as
Gibbs states for a one dimensional short range Ising
model\cite{[SRB],[R1],[G1]}.
\bibitem{[Bo]} The Eq(\ref{3.4}) as a rewriting the fluctuation
theorem Eq(\ref{3.3}) is remarkable and was pointed out by F. Bonetto.
\end{references}
\end{document}